\pdfoutput=1

\documentclass[sigplan, 10pt, nonacm=true]{acmart}

\renewcommand\footnotetextcopyrightpermission[1]{}
\newif\ifcomments

\usepackage{xspace}                         
\usepackage{subfig}                         
\usepackage[normalem]{ulem}                 
\usepackage{upgreek}
\usepackage{comment}                        
\usepackage{rotating}
\usepackage{graphicx}
\usepackage{multirow}

\newcommand{\printlayout}{{\bf The textwidth \the\textwidth; the columnwidth is \the\columnwidth}}

\newcommand{\eg}{e.g.,\xspace}
\newcommand{\ie}{i.e.,\xspace}
\newcommand{\egparen}[1]{(\eg~#1)\xspace}
\newcommand{\ieparen}[1]{(\ie~#1)\xspace}
\newcommand{\egp}[1]{\egparen{#1}}
\newcommand{\iep}[1]{\ieparen{#1}}

\newcommand{\myparagraph}[1]{\vspace{0.5em}\noindent {\bf #1:}}
 
\usepackage{hhline}
\usepackage{todonotes}
\usepackage{amsmath}
\usepackage{wasysym}
\usepackage{booktabs}
\usepackage{tabularx,ragged2e}
\usepackage{tabulary}
\usepackage{enumitem}
\usepackage{tikz}
\usepackage{listings}
\usepackage{mathdots}
\usepackage[outdir=./]{epstopdf}
\graphicspath{{./}}

\usepackage{enumitem}
\setlist{noitemsep, leftmargin=*, topsep=0pt, partopsep=0pt}

\newcommand{\ignore}[1]{}
\newcommand{\eat}[1]{}

\newcommand{\quickfig}[1]{Figure~\ref{#1}}

\newcommand{\tf}{TensorFlow\xspace}
\newcommand{\tfs}{TensorFlow's\xspace}
\newcommand{\tfz}{\tfs}
\newcommand{\system}{\textsc{PTF}\xspace}
\newcommand{\systemlong}{Pipelined \tf}
\newcommand{\systems}{\textsc{PTF}'s\xspace}
\newcommand{\systemz}{\systems}
\newcommand{\System}{\system}

\newcommand{\ptf}{\system}

\newcommand{\app}{\textsc{PTF}bio\xspace}
\newcommand{\appz}{\app's\xspace}

\newcommand{\gate}{gate\xspace}
\newcommand{\gates}{gates\xspace}
\newcommand{\gatez}{gate's\xspace}
\newcommand{\Gate}{Gate\xspace}
\newcommand{\Gates}{Gates\xspace}

\newcommand{\stage}{stage\xspace}
\newcommand{\stages}{stages\xspace}
\newcommand{\stagez}{stage's\xspace}
\newcommand{\Stage}{Stage\xspace}
\newcommand{\Stages}{Stages\xspace}

\newcommand{\alignsort}{align-sort\xspace}

\newcommand{\AlignSort}{Align-Sort\xspace}

\newcommand{\feed}{feed\xspace}
\newcommand{\feeds}{feeds\xspace}

\newcommand{\batch}{batch\xspace}
\newcommand{\batchs}{batches\xspace}
\newcommand{\batchz}{\batch's\xspace}

\newcommand{\clusterSize}{20\xspace}

\newcommand{\dataTPSIncrease}{4$\times$\xspace}
\newcommand{\dataMaxTP}{321 megabases/second\xspace}
\newcommand{\dataLatencyIncrease}{0.13$\times$\xspace}
\newcommand{\dataMaxTPrequiredOpenRequests}{6\xspace}

\newcommand{\slocCPP}{2476\xspace}
\newcommand{\slocPython}{841\xspace}

\newcommand{\dataMergeLatency}{298 seconds\xspace} 
\newcommand{\dataFusedVsBaselineDataSaving}{12\%\xspace}

\newcommand{\dataInputChunksPerRequest}{2236\xspace}
\newcommand{\dataSortsPerRequest}{224\xspace}
\newcommand{\dataGroupingFactor}{10\xspace}

\newcommand{\dataCCDFMeanRequestLatency}{420.6 seconds\xspace}
\newcommand{\dataCCDFTailRequestLatency}{479.7 seconds\xspace}

\newcommand{\dataCCDFMeanMergeLatency}{322.6 seconds\xspace}
\newcommand{\dataCCDFMeanAlignLatency}{2.7 seconds\xspace}
\newcommand{\dataCCDFMeanSortLatency}{2.7 seconds\xspace}

\begin{abstract}
  \tf is a popular cloud computing framework that targets machine learning applications.
  It separates the specification of application logic (in a dataflow graph) from the execution of the logic.
  \tfz native runtime executes the application with low overhead across a diverse set of hardware including CPUs, GPUs, and ASICs.
  Although the underlying dataflow engine supporting these features could be applied to computations beyond machine learning, certain design decisions limit this broader application, such as the inability for an application to differentiate between data items across concurrent requests.

  This paper introduces \systemlong (\system), a system that extends \tfz semantics to provide support for a broader variety of application logic.
  In particular, \system supports applications that concurrently process finite batches of data on a single instantiation.
  \ptf adds these semantics by partitioning the dataflow graph into a pipeline of smaller graphs and tagging each data item with metadata.
  These smaller graphs are separated by \emph{\gates}: new data structures in \ptf that buffer data items between graphs and interpret the metadata to apply the new semantics.
  \systemz pipeline architecture executes on an unmodified \tf runtime, maintaining compatibility with many existing \tf library functions.
  Our evaluation shows that the pipelining mechanism of \system can increase the throughput of a bioinformatics application by \dataTPSIncrease while only increasing its latency by \dataLatencyIncrease.
  This results in a sustained genome alignment and sorting rate of \dataMaxTP, using the compute and I/O resources of \clusterSize computers.
\end{abstract}

\date{}

\title{Extending \tfz Semantics with Pipelined Execution
}

\author{Sam Whitlock}
\affiliation{EPFL, Switzerland}
\email{sam.whitlock@epfl.ch}
\author{James Larus}
\affiliation{EPFL, Switzerland}
\email{james.larus@epfl.ch}
\author{Edouard Bugnion}
\affiliation{EPFL, Switzerland}
\email{edouard.bugnion@epfl.ch}

\begin{document}

\maketitle

\section{Introduction}
\label{sec:intro}

As the resolution with which scientific tools can measure experiments increases, so, too, does the amount of data they produce.
Diverse fields such as particle physics~\cite{cern-data}, bioinformatics genome sequencing~\cite{genome-moore,amplab-adam}, medical imaging~\cite{spark-mri,spark-fmri}, aeronautics~\cite{airbus-sensors}, and IoT produce data with a growth rate that exceeds Moore's law.
Users typically process such data sets with cloud computing frameworks that separate application logic from execution, delegating the latter to a common runtime deployed across large clusters of machines~\cite{bigdata-bench,bigdata-bench-one}.
These frameworks expose abstractions such as basic map-reduce~\cite{mapreduce}, but more general-purpose frameworks~\cite{dryad,flume,spark,naiad} encapsulate application logic in a directed acyclic graph (DAG), with each node expressing an operation \egp{reading a file} and the edges expressing data dependencies between nodes.
Each framework's runtime is responsible for
(a) instantiating the operations and stateful elements contained in the DAG across a set of scale-out hardware resources
and
(b) managing the execution of the application according to the logic corresponding to the DAG.

The specific type of framework typically used in big data scientific processing are batch frameworks.
Each request to the framework \egp{from a client program or work queue} corresponds to a finite number of data items, \eg the locations of all files to read and process.
Applications written with batch frameworks are instantiated once and process an unending stream of requests.

Cloud computing frameworks employ the \emph{pipelining} concurrency strategy to more effectively utilize a cluster's hardware resources.
Frameworks typically pipeline the data items in a request across the nodes in the application's DAG;
based on the semantics of the framework and the logic in the application DAG, each node in the DAG may concurrently process different data items from a request.
This increases the overlapping of computation with I/O, which increases hardware resource utilization.
Frameworks may pipeline requests as well, thereby overlapping the concurrent processing of data items from multiple requests.

\tf is a popular framework built on the \emph{dataflow} abstraction that targets machine-learning applications.
\tf encodes application logic using a DAG and includes a large library of nodes for mathematical operations;
it is a popular choice in the machine learning domain, with applications to cancer research, medical imaging, retinal disease, and physics~\cite{tf-skin-cancer,tf-gans,tf-medical-images,tf-retinal-disease,tf-physics}.
\tfz highly optimized runtime instantiates application graphs across multiple machines, places each node on a machine, and inserts the necessary communication mechanisms between nodes on different machines or to a hardware accelerator.
The runtime is written in C++ to take advantage of heterogeneous hardware, including CPU vector units, GPUs, and even custom ASICs~\cite{tpu}.

The performance of \tfz native execution provides an ideal basis for a cloud computing framework for scientific applications as well as machine learning applications.
Contemporary JVM-based cloud computing frameworks can impose a 16-43$\times$ overhead for analytics workloads as compared to a framework based on C++~\cite{nimbus-performance}.
The JVM~\cite{jvm} itself imposes overheads for of 1.9-3.7$\times$ compared to a native language for computationally-intensive applications~\cite{hundt2011loop,jvm-cpp-3d}.
Scientific applications are heavily impacted by such overheads due to their high ratio of cycles per byte of I/O~\cite{cycles-per-byte-science,big-data-bio}.

Unfortunately, a key design decision limits \tfz applicability as a general-purpose cloud computing framework: 
\tf cannot pipeline multiple requests within the runtime.
Although \tf can pipeline data items of a single request, it cannot distinguish data items associated with different user requests.
While workarounds exist \egp{performing the disambiguation in the client code controlling \tf via the Python API}, this leads to expensive data conversion~\cite{weld}.
This limitation is acceptable in \tfz target domain of machine learning, which typically runs applications with a single coarse-grain graph representing a long-running computation.


This paper presents \emph{\systemlong} (\system), a system that bridges this gap between \tf and cloud computing frameworks.
\system enables the continuous execution of concurrent, multi-request pipelines --- a key functionality for cloud computing frameworks --- within the \tf runtime.
\system applications can overlap the parallel operation of application I/O and compute phases.
\system adds two key properties to \tf:

\myparagraph{Concurrent and isolated execution} 
\system provides the abstraction of an isolated pipeline for each request.
Regardless of when the request is submitted to an application or the number of concurrent requests, \system ensures that each request is processed as if it were the only request submitted to the application.
 
\myparagraph{Bounded resource utilization}
\system implements a two-level, credit-based flow control within an application's pipeline to ensure bounded resource utilization when concurrently processing a stream of requests.
This is crucial for memory-intensive, scientific applications, which should operate without swapping, for efficiency.
The two-level scheme separates flow-control considerations within and across machines.

\system adds three abstractions to \tf that enable multi-request pipelining.
(1) \emph{\Stages} encapsulate \tf graphs representing a small subcomponent of the application's logic.
(2) \emph{\Gates} separate adjacent \stages and use additional metadata to differentiate inputs between concurrent requests.
(3) \emph{Pipelines} are composed of a series of \gates and \stages.
\system implements these abstractions entirely within the existing \tf runtime as a small additional library of code in the \tf codebase.

Our contributions include:
\begin{itemize}

\item Three new abstractions (\stages, \gates, and pipelines) that extend \tfz semantics and capabilities to enable multi-request pipelining 

\item \system, a backward-compatible patch to \tf that implements these new abstractions 

\item \app, a high-throughput, pipelined bioinformatics application service.
  \app integrates the Persona~\cite{persona} data format and aligners into \system.
\end{itemize}

We evaluate \app to characterize \systemz scale-out behavior on a cluster of \clusterSize servers.
We show that
(a) the use of pipelining increases throughput by \dataTPSIncrease while only increasing the average latency of processing an individual request by \dataLatencyIncrease
and
(b) linear speedup to the size of our cluster.

\system is open-source and can be found in the following repositories:
\begin{itemize}
\item \texttt{\url{https://github.com/epfl-dcsl/ptf-system}}, which contains the code for \system and Persona as additions to a standard \tf code repository.
\item \texttt{\url{https://github.com/epfl-dcsl/ptf-persona}}, which contains the Python code that constructs and executes local and scale-out applications using \system.
\end{itemize}
We refer the reader to the documentation in these repositories for further details.
 
The rest of the paper is organized as follows: \S\ref{sec:background} provides the necessary background on cloud computing frameworks and \tf.
We then discuss the architecture (\S\ref{sec:arch}) and implementation (\S\ref{sec:implementation}) of \system.
In \S\ref{sec:persona}, we describe the construction of \app, including a novel fused alignment/sort step in the pipeline.
We evaluate the performance of \system (\S\ref{sec:eval}), discuss related work (\S\ref{sec:related_work}) and conclude (\S\ref{sec:conclusion}). 

\section{Background}
\label{sec:background}

This section describes the paradigm of contemporary cloud computing and streaming frameworks, \tfz dataflow model, and how they differ.

\subsection{Cloud Computing Frameworks}
\label{sec:background_streaming}

Cloud computing frameworks~\cite{google-dataflow,flink,spark-streaming,dryad,heron-twitter,samza,amazon-kinesis,naiad,streamscope,spark,mapreduce,flume,storm-twitter,millwheel,nimbus} are a popular architecture to orchestrate data processing in data centers because they separate application logic from execution.
Application logic is typically encoded as a directed graph of data dependencies between successive nodes, each of which performs a modular subcomponent of the application's overall logic \egp{reading a file, summing all input values}.
This graph is typically acyclic, or contains a few coarse-grained loops, \eg a global loop to train a classifier.
The framework's runtime executes the application by launching a set of tasks \egp{threads, processes, containers} on the underlying hardware that each execute one or more kernels corresponding to nodes in the graph.
The runtime enables application developers to focus on the application logic, not on the low-level details of how the logic is implemented;
the graph is not a procedural specification of the computation.

Cloud computing frameworks increase resource utilization by concurrently processing requests.
An application written using a cloud computing framework is typically allocated a fixed set of resources upon instantiation by a resource manager~\cite{yarn,kubernetes,mesos,borg};
each request submitted to the framework uses a subset of these resources.
As the framework processes a request, it must
(a) adapt the subset of resources allocated to each request based on all request-associated allocations \egp{delaying requests to avoid over-allocation}
and
(b) provide the semantics of an isolated set of resources to each request so that
the result of a request is not affected by concurrently-executing requests.

Delegating resource partitioning to a cloud computing framework enables fine-grained resource partitioning schemes that maximize resource utilization.
Permanently assigning hardware resources to a request wastes resources because the resource manager has little insight into the application logic.
The associated overhead is due to:
(a) resource stranding within an application \egp{allocation of an expensive resource that cannot be shared between requests},
(b) the necessity to run multiple copies of the same application logic (one per request),
and
(c) the constant allocation and deallocation of shared resources.
By using application logic to determine when and how to safely share and partition resources, cloud computing frameworks can multiplex concurrent request processing at a finer granularity.

Data pipelining is a common strategy used to achieve this finer-granularity resource sharing.
Pipelining is the concurrent processing of distinct data items by different operations in the application logic, \ie corresponding to different nodes in the DAG.
These data items may be associated with the same request or different requests, depending on the capabilities of and semantics supported by the framework.
By overlapping different computations in the application, the cloud computing framework's runtime may use a greater portion of the underlying hardware resources at any given time, \eg overlapping a node performing I/O with another doing computation.
This pipelining strategy also increases the likelihood that any given node in the application's DAG will have an input ready when it finishes processing a data item, thereby spending less time waiting for input from upstream nodes in the application DAG.

\subsection{\tf Programming Model}
\label{sec:background:tf}

\tf uses dataflow to express application logic as operations on tensors.
Dataflow represents application logic as a DAG: nodes represent operations and edges specify the propagation of the results.
\tfz features and design decisions are based in its focus on the machine learning domain; most of its predefined nodes perform stateless numeric computations.
Each node consumes and emits only tensors, multi-dimensional arrays of a single elementary data type (numerical or string).

\tf graphs process tensors using a push model of execution.
A feed is a group of tensor values that populate placeholder inputs to a graph.
The client program inputs a feed into a graph and requests the value of the graph's output, \ie a downstream group of tensors.
The \tf runtime responds to this request by propagating the feed through the graph based on a strict set of rules~\cite{tf-model,tagged-dataflow}.
The result is a new feed (corresponding to the downstream tensors) and side effects, \eg updating a stateful variable held by a graph node.

A batch of feeds is a collection of related feeds that together form the input for an application, \eg a series of labeled images on which to train a machine learning model~\cite{mnist}.
Upon startup, \tf applications initialize stateful variables \egp{large tensors that hold persistent data across successive feeds} and then process a batch of feeds.
The client program drives each feed through the \tf runtime in succession, updating the variables \eat{holding model parameters} each time.
A single graph may process concurrent feeds from a single request, but extra information must often be added to the graph in order to serialize parallel updates to variables.

Monolithic application graphs can be decomposed into smaller graphs separated by \tf queues.
Each \tf queue separates successive phases of an application into distinct independent graphs;
these graphs are linked by the \tf queue data structure that buffers feeds between upstream and downstream graphs.
This separation increases concurrency within an application.
For example, an initial phase typically reads in a batch of feeds from storage (training examples) and a subsequent phase trains a model based on the values.
Each graph receives a feed from its upstream queue via a dequeue node and sends the resulting feed to its downstream queue via an enqueue node.
Each graph is driven by a queue runner, \ie a Python thread containing no application logic that drives feeds through a graph by requesting the value of the graph's enqueue node.

The \tf framework targets applications that process a single batch of feeds per invocation.
While internal concurrency between feeds is possible, \tf does not natively distinguish between feeds belonging to different batches of inputs.
A multi-batch \tf application must rely on the client program to disambiguate between feeds from different batches.
The necessary performance penalty for involving the client program is the copying of data into and out from the \tf runtime.
Machine-learning \tf applications are not inhibited by this design decision, as they typically do not concurrently process multiple batches;
their graphs contain computationally-intensive coarse-grained operations that have little overhead to construct on a per-batch basis.

\section{Architecture}
\label{sec:arch}

\system is a cloud computing framework built on \tf that allows for concurrent, isolated, and flow-controlled processing of a stream of successive requests.
These requests correspond to \batchs of \feeds and are supplied from a common single queue.
\system dequeues requests from the queue in succession and processes them to completion entirely within the \tf runtime.

\system is a careful addition of code to \tf to enable pipelines of \tf graphs to process concurrent \batchs within the same invocation of an application.
This pipeline of independent \tf graphs enable \system to support data pipelining: all graphs in the pipeline can simultaneously process different data items (\feeds).
This architecture enables semantics necessary for \system to operate as a more general-purpose cloud computing framework, performing both data processing and \batch management \iep{tracking the progress of each \batch in a pipeline};
it does not rely on an external framework or client-side code \iep{code calling into the \tf runtime via its API} in order to perform any of its operations, thereby avoiding expensive data copying.

The key insight of \system is the association of an additional metadata tensor with a \feed.
The metadata tensor embeds the necessary information to
(a) identify the \batch to which a \feed is associated
and
(b) describe the \batch.
\system interprets the metadata directly in the \tf runtime.
Passing this metadata as a tensor within a \feed enables \system to reuse \tfz runtime to pass management information with data.

\system inherits key attributes from \tf: (1) it allows for the concurrent execution of an arbitrary graph across multiple cores and heterogeneous hardware components;
(2) it scales out across multiple machines with negligible overheads by forgoing the need for any global coordination when partitioning the workload across machines.

\subsection{Components}
\label{sec:arch:components}

\begin{figure}[!t]
    \centering
    \includegraphics[width=\columnwidth]{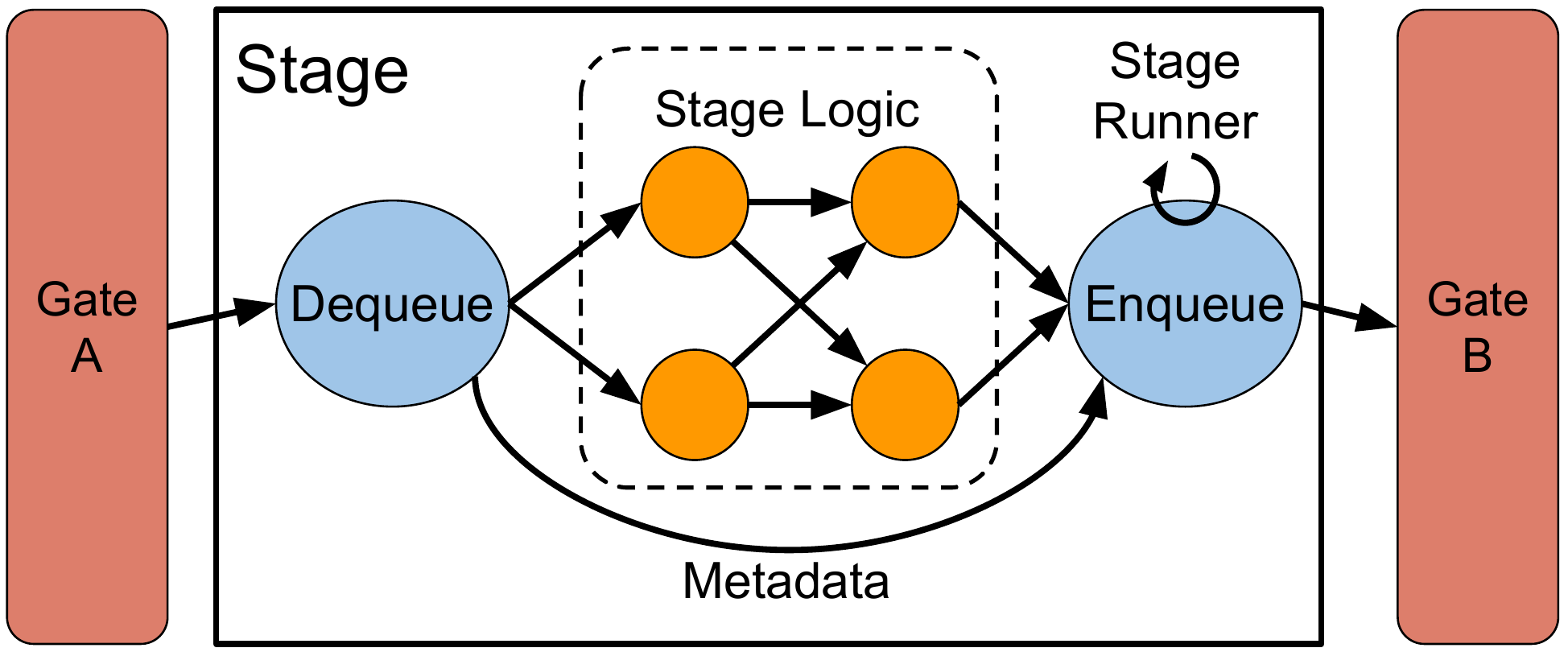}
    \caption{The components of a \stage: the logic, the adjacent \gates with the corresponding enqueue and dequeue nodes in the \stage, and the metadata.}
    \label{fig:stage}
    \label{fig:stage_detail}
\end{figure}

\system expresses applications as a pipeline of successive \stages synchronized via connecting \gates.
Each \stage is a \tf graph that statelessly transforms \feeds to apply the subcomponent of the application logic it represents.
The \gates synchronize and coordinate the concurrent execution of different \feeds throughout the pipeline.
This decomposition of a large \tf graph into smaller graphs (\stages) decouples the large chain of data dependencies between input and output, enabling \systemz \gates to interpose between different \stages to apply its new functionality.

\quickfig{fig:stage_detail} shows the details of a \stage.
The \tf graph in the \stage interacts with adjoining \gates via enqueue and dequeue nodes to send and receive \feeds, respectively.
A \stage runner thread drives the \stagez graph with successive invocations via the Python API, repeatedly checking the upstream \gate to trigger the \tf runtime to execute the \stage on new \feeds;
\stage runners serve a similar purpose as queue runners for \tf queues, \ie driving the execution of a \stage without performing any logic in the Python code.
The metadata tensor from the \feed is passed around the core \tf graph, as the application logic of the graph does not alter the metadata tensor.

\Gates store \feeds in a buffer between successive \stages so that a pipeline may execute multiple \feeds concurrently.
Similar to \tf queues, \gates interpose between adjacent separate graphs to decouple the direct data dependencies between two successive \stages.
\Gates use this decoupling to apply their custom concurrency and pipelining logic, as they are able to regroup, reorder, or delay the \feeds without usurping \tfz dataflow rules.

\Gates coordinate data availability and buffering between adjacent \stages.
Once a \stage processes all tensors in a \feed, it atomically inserts the entire \feed into its downstream \gate.
The \gate buffers the \feed until the downstream \stage requests a new output from the \gate.
The \gate then selects a \feed from its buffer and atomically transmits all \feed elements to the downstream \stage for processing.
The execution of each \stage is limited by the availability of
a) a \feed from the upstream \gate and
b) capacity to enqueue the resulting \feed into the downstream \gate.
Any \stage that has both may execute concurrently with any other \stage in the pipeline.

\Gates interpret the metadata contained in a \feed to apply \systemz concurrency and pipelining semantics.
When a \feed is enqueued into a \gate, it is stored and dequeued based on its associated \batch.
As concurrent \batchs progress through a pipeline, each \gate stores and forwards the constituent \feeds such that the result of processing any given \batch is identical to processing the \batch in a non-multiplexed pipeline.
This isolated pipeline abstraction is enabled by the custom logic in the \gates and the stateless nature of \stages.
The custom logic is performed directly in the \tf runtime so that \system does not rely upon an external framework or client program.

The metadata consists of two integers: an ID and an arity.
The ID is a unique numerical identifier assigned to a \batch when it enters a \system pipeline.
The arity is the number of \feeds in the \batch.
By construction, all \feeds from the same \batch will have identical metadata.

\subsection{\Gate Operation}
\label{sec:arch:gates}

\Gate operations update the state of the \gate by inserting or retrieving a \feed with its associated metadata.
A \stagez graph first dequeues a \feed from the upstream \gate via a dequeue node. 
Once that \feed has propagated through the \stagez graph, it terminates in an enqueue node, which inserts the \feed into downstream \gate.
Both the enqueue and dequeue operations are synchronous: further computation in the \stage blocks until their respective operations complete.

\Gates serve enqueue and dequeue operations from a single operation queue, protecting concurrent access to its internal data structures with locks.
A \gate may reorder operations in the queue based on its operational semantics.
For example, an enqueue operation will be served before any dequeue operation if there are no buffered \feeds.
In the common case, these operations are performed in a first-come, first-serve (FCFS) order.

\Gates interpret the metadata to track and control the progress of \batchs through the application pipeline.
The \gate examines the ID of a \feed to determine if the \feed is from an existing \batch.
If the \feed is associated with a new \batch, the \gate allocates space in its \feed buffer for the \batch and inserts the new \feed.
If the \feed belongs to an existing \batch, the \gate inserts the \feed into the preallocated \feed buffer space for that \batch.

\Gates control the lifecycle of a \batch by deciding when to open and close each \batch it processes.
A \gate opens a new \batch by beginning to send \feeds from the buffer.
Dequeue operations remove \feeds from the buffer until the \gate determines that no more \feeds will be available by examining the metadata arity and the number of \feeds buffered and dequeued.
The \gate closes the \batch by freeing the associated data structures, including the space in its \feed buffer.
A \batch may be opened before all of its \feeds have been enqueued into the \gate.

\Gates may emit \feeds in any order from any open request.
When selecting which \feed to emit to the downstream \stage through a dequeue operation, a \gate may choose any \feed from any request that is open and has \feeds buffered.
This loose ordering guarantee enables \system to improve concurrency and \feed pipelining by increasing the number of \feeds buffered in the \gate that are candidates to emit downstream.
In practice, \gates emit \feeds in preferential order in which \batchs are opened;
\feeds are emitted in FIFO order within a given \batch, based on their arrival into the \gate via an enqueue operation.

\Gates use special variants of enqueue and dequeue operations to consume and produce aggregate \feeds.
\Gates combine multiple individual \feeds from a \batch to produce an aggregate \feed, which contains the same number and type of tensors as the original \feed type, but with an additional dimension added to each tensor in order to group the individual \feeds' tensors \egp{a vector to a matrix}.
Aggregate dequeue operations change the arity because the reduce the total number of \feeds in the \batch;
the downstream \stage consuming an aggregate \feed must produce a single output \feed, thereby reducing the arity by the aggregation size.
If the requested aggregate size is $S$ and the original arity is $A$, then the new arity is calculated as $\lceil A \div S \rceil$.
All of the \feeds in each aggregate \feed produced by the aggregate dequeue operation originate from the same \batch and contain $S$ individual \feeds, except for the last aggregate \feed produced for the \batch that may contain $A\bmod S$ \feeds if that value is greater than 0.

Aggregate dequeue operations are used for a variety of purpose in a \system pipeline.
They can be used to provide multiple input \feeds at once to a \stage, which may enable more efficient nodes to be used in the \stagez graph.
An aggregate dequeue operation can serve as a barrier in a \system pipeline by aggregating all items in a \batch before emitting them downstream; in this case, the requested aggregate size must be greater than any \batchz arity.

\subsection{Resource Bounding}
\label{sec:arch:bounding}

\System provides credit-based flow control between any two \gates.
Each credit in \system represents the ability of a \gate to open a new \batch.
Credits are issued by a downstream \gate to the linked upstream \gate.
When the upstream \gate receives a credit, it may open a new \batch and begin to send the associated \feeds.
When a \gate closes a \batch, it increments the number of available credits, sending them to the upstream \gate if one is linked.

\Gates can locally limit the size of their \feed buffer to control resource usage regardless of \batch size.
A \gate with a non-aggregate dequeue may limit the total size of its \feed buffer for all open \batchs.
When the \feed buffer is full, enqueue operations block until subsequent dequeue operations free space.
This resource bounding mechanism is similar to that which is used by \tf queues to bound resource usage.

\subsection{\Stage Parallelism}

A \system application can replicate any \stage to scale out across local hardware resources.
Replicating a \stage exposes more nodes to the \tf runtime, which increases the degree of parallelism subject to \feed availability from the upstream \gate.
A \stage is replicated by copying its \tf graph, \stage runner, and inserting new enqueue and dequeue nodes for each replica.
Each replicated \stage operates independently of other \stage replicas; it executes only when the upstream \gate sends an input \feed to it.
The upstream \gate serves \stage replica requests in FCFS order.

\eat{
This scaling strategy relies on the stateless nature of each \stagez graph.
Each \stage is agnostic about which subset of the \feeds of any given \batch it may receive, if any.
}

\subsection{Pipeline Hierarchy}

\System scales out pipelines across multiple machines by using a two-level nesting hierarchy of \emph{local} and \emph{global} pipelines.
A local pipeline consists of \gates and \stages that are placed in a single process (typically one process per machine).
A global pipeline consists of a sequence of local pipelines separated by global \gates.
An application can replicate a local pipeline onto multiple machines to scale out across additional resources.
\Gates in the global pipeline coordinate progress of \batchs in the local pipelines.
Global \gates may be placed on any machine in the cluster; by default, \system places them on a single dedicated machine to not disrupt local pipelines.

Global pipelines distribute \batch partitions to minimize communication overhead when distributing \feeds to local pipelines.
A partition is a subset of a \batch in the global pipeline that is distributed to a local pipeline.
A local pipeline processes each partition it receives as a standalone \batch.
Distributing partitions instead of \feeds from a global \batch avoids a scaling bottleneck by decoupling coarse-grained partition distribution from fine-grained \feed processing.

\Gates in the global pipeline create partitions by performing an aggregate dequeue operation and modifying the metadata to contain the ID and arity of both the original \batch and the partition.
Local pipeline \gates use only the partition metadata to perform their operations.
The subsequent global \gate reassembles the original \batch by stripping away the partition metadata and using only the \batch metadata.

\System bounds resources using a hierarchical flow control scheme.
Each \gate within a local pipeline may bound upstream \gates within the same local pipeline with a local credit link. 
\Gates in the global pipeline may bound upstream global \gates in order to limit global resource usage with a global credit link. 
Although credit linking in both the global and local pipelines does not provide a tight bound on the total resource usage (due to the fact that a given \batch may contain any number of \feeds),
it prevents \gates upstream of a thoughput-limiting \stage \egp{one that performs a computationally-intensive process} from accepting an unlimited number of \feeds in steady-state operation.
\Gates in either the local or global pipeline may limit the size of their \feed buffers.

\begin{figure*}[!t]
    \includegraphics[width=\textwidth]{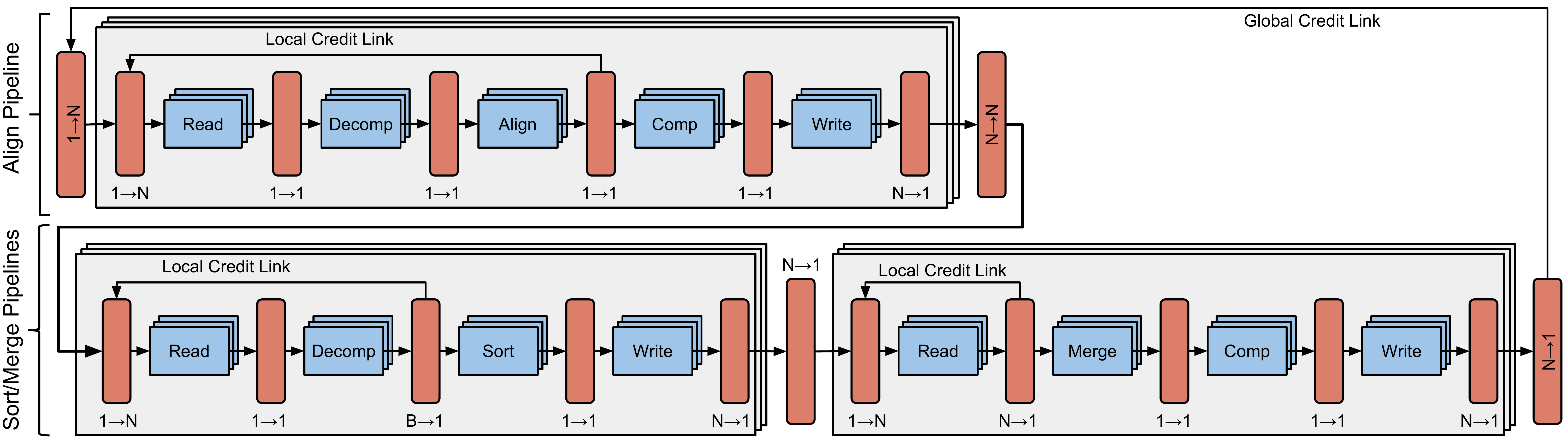}
    \caption{A diagram of the \emph{baseline} \system application containing 3 local pipelines (align, sort, and merge), each of which have multiple \stages that scale based on the underlying hardware.}
    \label{fig:persona_baseline}
\end{figure*}

\quickfig{fig:persona_baseline} shows an example of a \system application that scales across machines using global and local pipelines.
This shows the baseline application for \system Persona with three local pipelines: align, sort, and merge.
Pipeline and \stage replication are used to scale across global and local resources, respectively.
\Gates in both the global and local pipelines make use of aggregate dequeue operations: the sort \stage in the sort pipeline operates on aggregates of size $B$ ($B\rightarrow 1$) and the merge pipeline requests partitions containing the entire \batch of size $N$ ($N\rightarrow 1$).

\subsection{Computational Model}

\system does not change the standard \tf semantics~\cite{tf-model}; it partitions monolithic \tf graphs into smaller graphs \iep{\stages} in which the existing semantics apply.
By splitting the application into \stages, \system applies standard \tf semantics within each \stage and interposes between \stages to apply pipelining semantics with \gates.
This enables \system to use the \tf runtime with no modifications, including the reuse of the existing features in \tf such as
(1) the distributed runtime,
(2) the ability to serialize the entire application for distributing to each node and exporting, and
(3) the myriad existing \tf nodes.

\system is backward compatible with \tf, \ie any \tf node used to construct a \tf graph may be used to construct a \system graph.
Note, however, that a \tf program often consists of a graph combined with client logic, described in \S\ref{sec:background:tf}, which typically iterates over \feeds until some termination condition is met.
\system, with its design goal for cloud computing applications, does not support client-driven logic not expressed in the graph.
Instead, \system submits each \feed exactly once to the graph and any iterative or conditional construction must be expressed within the graph itself.

\system uses the dataflow semantics of \tf to avoid using a centralized scheduler.
When a \stage dequeues a \feed and its associated metadata, \tfz guarantees that the \stagez graph emits exactly one resulting \feed after processing the input \feed.
This guarantee ensures that the metadata remains a valid description for all \feeds in any given \batch because neither the ID nor the arity can be invalidated by unpredictable graph behavior \egp{emitting more than one result \feed for a given input \feed}.
\tfz exactly-once delivery semantics ensures that each \gate only receives a given \feed once.
\Gates rely on this guarantee to track the progress of concurrent \batchs without a centralized or broadcast-based scheduling architecture;
the metadata provides sufficient information for any \gate to locally track the progress of a \batch based solely on \feed arrival, regardless of the complexity of the upstream and downstream \stage graphs.

\section{Implementation}
\label{sec:implementation}
\label{sec:impl}

The \system patch to \tf consists of \slocCPP lines of C++ code and \slocPython lines of Python. 
Apart from the build system, \system does not require modifications to the \tf runtime or existing library of nodes.
The patch defines \gates, \stages, and pipelines, which are exposed via the Python API to enable a user to construct a \system application.

The core \system API defines \gates and \gate operations.
\Gates are instantiated as shared resources within the \tf runtime.
The enqueue and dequeue operations that manipulate a \gatez internal data structures are colocated in the same process as the \gate.

\System applications utilize \tfz distributed runtime to coordinate global and local pipeline execution.
An application first creates the complete \tf graph describing all \gates, \stages, and pipelines.
The nodes that define the operations and resources in this graph are each annotated with a logical device \iep{a label in the \tf graph attached to each node}.
The complete graph is then distributed to all machines in the cluster.
Each machine is assigned a logical device, corresponding to a local pipeline;
upon startup, the machine instantiates the nodes and \stage runners corresponding to its device and connects to other machines in the cluster via the network.
Once the startup process is complete, the application serves client requests via a request-response mechanism \egp{a web server}.

\section{Using \system for Bioinformatics Applications}
\label{sec:persona}

\begin{figure*}[!t]
    \centering
    \includegraphics[width=\textwidth]{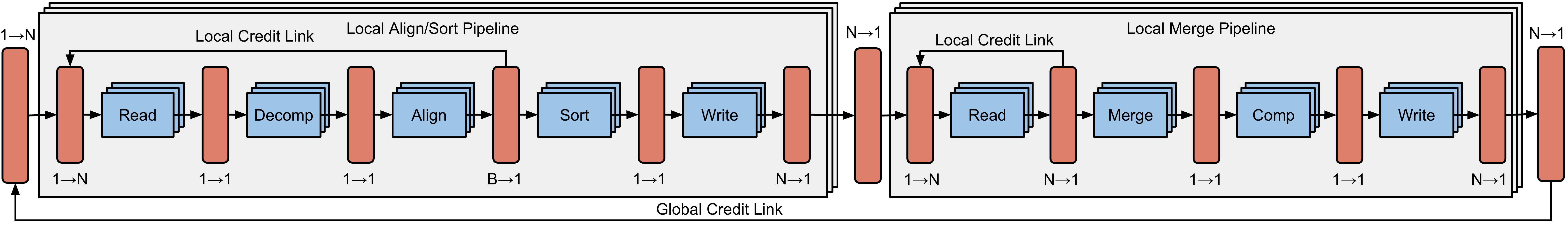}
    \caption{A full diagram of the Persona \AlignSort \system pipeline, with the align and sort phases fused into a single pipeline.}
    \label{fig:persona_fused}
\end{figure*}

We demonstrate the features of \system through the construction of \app, a distributed bioinformatics framework and set of applications.
\app combines \system with Persona~\cite{persona}, a framework written in \tf that enables users to construct scale-out pipelines for bioinformatics computations ranging from genomic sequencing to protein matching.
Persona leverages \tfz native code execution to incorporate popular bioinformatics applications such as SNAP~\cite{snap} and uses the Aggregate Genomic Data (AGD) data format, a new chunked column-oriented data format designed that enables scale-out processing of large genomic and proteomic datasets.

\appz use of \system enables the construction and deployment of persistent, flexible bioinformatics pipelines as a service.
While Persona benefits from the use of \tfz many features \egp{flexible graph composition, queues, schedulers}, \tfz semantics limited it to processing a single user request per invocation.
This per-request instantiation prevented high startup costs \egp{memory mapping large files and warming up buffer pools} from being amortized across multiple requests.
To demonstrate the flexibility of the approach, we focus on a genomic application that:
\begin{itemize}
\item Aligns the reads of a genomic dataset captured by a sequencing device against the reference human genome.
  This is a computationally-intensive process that leverages SNAP to first determine candidate locations, compute the edit distance, and determine the candidate location(s).
\item Sorts the resulting locations against the human genome itself, implemented as an out-of-core merge sort.
  The merge sort has a parallel sort phase, which writes out intermediate sorted files, and then a serial merge phase, which reads in all intermediate files to produce a sorted dataset.
\end{itemize}
Although some datasets may fit in the memory of a single machine, an out-of-core sort enables higher throughput because the sort phase may run in parallel across multiple machines.
\eat{
an out-of-core merge sort is advantageous because the compression associated with this phase (decompression when reading and compression when writing out the results) becomes a bottleneck;
it enables compute-bound compression phases to be spread across multiple machines by utilizing spare network capacity.
}

We implement two variants of this pipeline in \app.
First, the baseline application, shown in \quickfig{fig:persona_baseline}, connects three serial pipelines together: first to align the individual reads (which generates an additional AGD column with the alignment information), then separately to sort the AGD chunks in large batches, and finally to perform the final merge stage to generate the full genomic dataset aligned and sorted against the reference genome.
Each pipeline consists of multiple stages to overlap I/O phases with computational phases.
It is a persistent service that is analogous to Persona's solution of executing successive, distinct batch jobs to first align and subsequently sort a dataset using a two-phase out-of-core merge sort.

\quickfig{fig:persona_fused} illustrates the enhanced solution, which fuses the alignment step with the sorting step:
rather than writing the alignment column to disk and then sorting, the fused \alignsort phase aligns and sorts partitions in the same local pipeline.
The sort \stage uses an aggregate dequeue operation to sort up to $B$ AGD chunks in memory and write out the sorted result (a single, aggregate AGD chunk per $B$ input chunks) to disk.
In effect, the fused \alignsort uses spare memory capacity and NIC bandwidth on the alignment machines to eliminate one full I/O read and write cycle for the dataset in addition to the associated CPU resources for compression in the baseline version.

Both variants of the pipeline limit the number of open \batchs in the global and local pipelines via global and local credit links.
The global credit links are end-to-end, limiting the total number of open \batchs in the pipeline at any given time.
Local credit links bound memory usage of a local pipeline.
For example, the local credit link in the merge pipeline of \quickfig{fig:persona_fused} prevents the read \stages from reading more chunk files than the merge \stages can handle, thereby limiting the amount of memory consumed by buffers receiving the data from the storage system.

\quickfig{fig:persona_fused} also illustrates how \app uses \systemz abstractions and mechanisms: 
(1) The application consists of two serially-connected pipelines, each of which has a distinct, configurable, scale-out width to take advantage of the scale-out cluster. 
(2) The use of metadata tags allows for the concurrent, isolated execution of multiple \batchs.
For example, the inherently serial dependency between the two pipelines and the serial nature of the merge phase of the out-of-core merge sort provide a key opportunity for throughput improvements without any impact on latency.
(3) The use of credit-based flow-control is used both for admission control at the front of the pipeline (to control the degree of \batch concurrency) as well as within the pipeline (to control the amount of memory consumed by the AGD chunks as they flow through the pipeline).
(4) Both pipelines show a typical pattern of a central computational phase (alignment, sorting, or merging) surrounded by a read/decompress phase and a compress/writeback phase.

\section{Evaluation}
\label{sec:eval}

We next evaluate \systemz ability to operate as a cloud computing framework through an evaluation of \app.
Specifically, we evaluate how it achieves the architectural goals outlined in \S\ref{sec:intro} by evaluating the fused \alignsort system described in \S\ref{sec:persona}.
To do so, we begin by evaluating the performance benefit of pipelining requests in the application.
Second, we evaluate the scale-out performance of the application as we increase the cluster size.
Finally, we evaluate the benefits of fusing the align and sort pipelines and \appz ability to overlap I/O with computation.

Our evaluation shows that the semantics provided by \system enable persistent streaming big-data applications, specifically:  
\begin{itemize}
\item \system enables pipelining of parallel \batchs on the same pipeline until a component of the pipeline becomes saturated.
\item \system scales out across multiple machines to the point of hardware saturation.
\item \system overlaps I/O and compute to enable efficient resource utilization
\end{itemize}

\subsection{Experimental Setup}
\label{sec:experiment_setup}

We use a cluster of \clusterSize typical datacenter machines, each with two Intel Xeon E5-2680v3 CPUs at 2.5GHz, 256 gigabytes of DRAM, and a 10GbE network interface.
We enable hyperthreading on all 12 cores per socket, for a total 48 logical cores per machine.
All machines run Ubuntu 18.04 Linux with the distribution's default Linux kernel (4.15.0).
The compute and storage are connected by a 40GbE-based IP fabric consisting of 8 top-of-rack switches and 3 spine switches.

The benchmarks read and write to a genomic database of paired-end whole human genome dataset from Illumina~\cite{dataset} (Platinum dataset, ERR174324), which consists of 223 million single-end 101-base reads. 
Input and output files are stored in the Persona AGD format~\cite{persona} with a chunk size of 100,000 \iep{the number of records in each chunk file}.
We store the input datasets, intermediate files, and output datasets on a Ceph distributed object store~\cite{ceph} comprised of 18 nodes, each with 10 disks.
\app accesses Ceph objects via the RADOS API\@.
\eat{
The highest-throughput configuration of the fused \alignsort application reads \dataFusedCephRead and writes \dataFusedCephWrite during a period of \dataFusedCephPeriod.
}

All experiments test the application running as a service processing parallel client requests.
Each request is comprised of a list of keys corresponding to the AGD chunk files for a dataset.
Each dataset represents a single-end dataset of one entire individual from the Illumina Platinum dataset.

The latency for each request is defined as the service time of a request once it is submitted to the pipeline.
The throughput is measured as the number of bases (in millions, \ie \emph{megabases}) processed per second.

All local pipelines configure a sufficient number of \stages such that a local resource is saturated (CPU, NIC, or main memory).
Any additional \stages do not increase performance, as all \stages are executed by the \tf runtime using a fixed-size thread pool per machine.
Each additional \stage replica specifies the maximum possible parallelism for a given \stage, but the \tf runtime decides which nodes amongst all \stages to execute based on \feed availability.

\subsection{Benefits of Pipelining}
\label{sec:eval:bounding}

\begin{figure}[!t]
    \centering
    \includegraphics[width=\columnwidth]{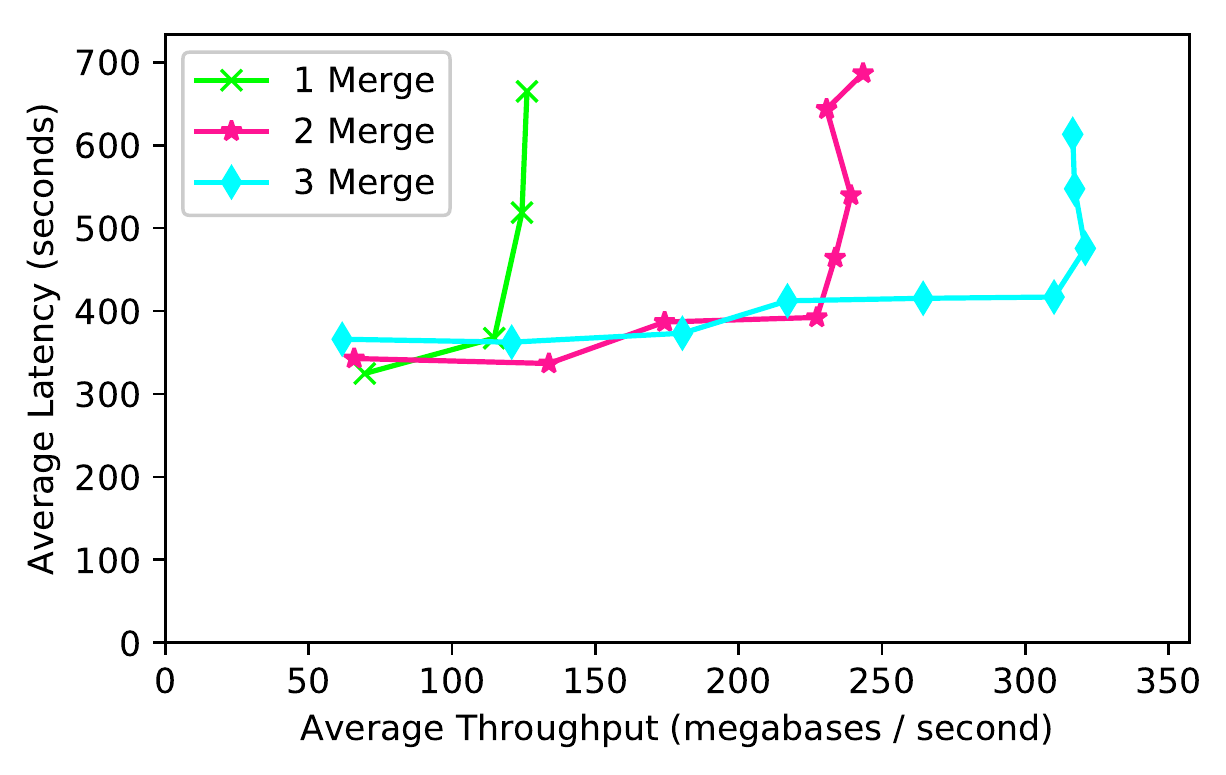}
    \caption{Latency vs. throughput for the fused \alignsort application. Each series has a fixed number of local merge pipelines (1 to 3) and \alignsort pipelines.
      Each series shows an increasing number of open \batchs configured on the same application, beginning from a single \batch.
      }
    \label{fig:latency_vs_throughput}
\end{figure}

\quickfig{fig:latency_vs_throughput} shows the effects of the number of open \batchs on the throughput and latency for the fused \alignsort application.
The results show that each additional open \batch increases overall throughput until a hardware component becomes saturated.
After this point of saturation, additional open \batchs must queue in the buffer of the upstream \gate to await processing.

This result validates the benefits of request \iep{\batch} concurrency and \feed pipelining.
\system is able to coordinate the additional parallelism of processing \feeds from multiple concurrent \batchs to better utilize the cluster's hardware resources.
This results in additional throughput with negligible or no latency increase until one component becomes saturated \iep{the merge pipelines}.
\system is able to maintain this post-saturation throughput by buffering \batchs at the global pipeline level.

With \dataMaxTPrequiredOpenRequests open \batchs, the configuration with 3 merge pipelines is able to achieve \dataMaxTP in its maximal configuration \iep{17 fused \alignsort pipelines}, an increase of \dataTPSIncrease over the 1 fused \alignsort pipeline configuration with a \dataLatencyIncrease increase in request latency.

\begin{figure}[!t]
    \centering
    \includegraphics[width=\columnwidth]{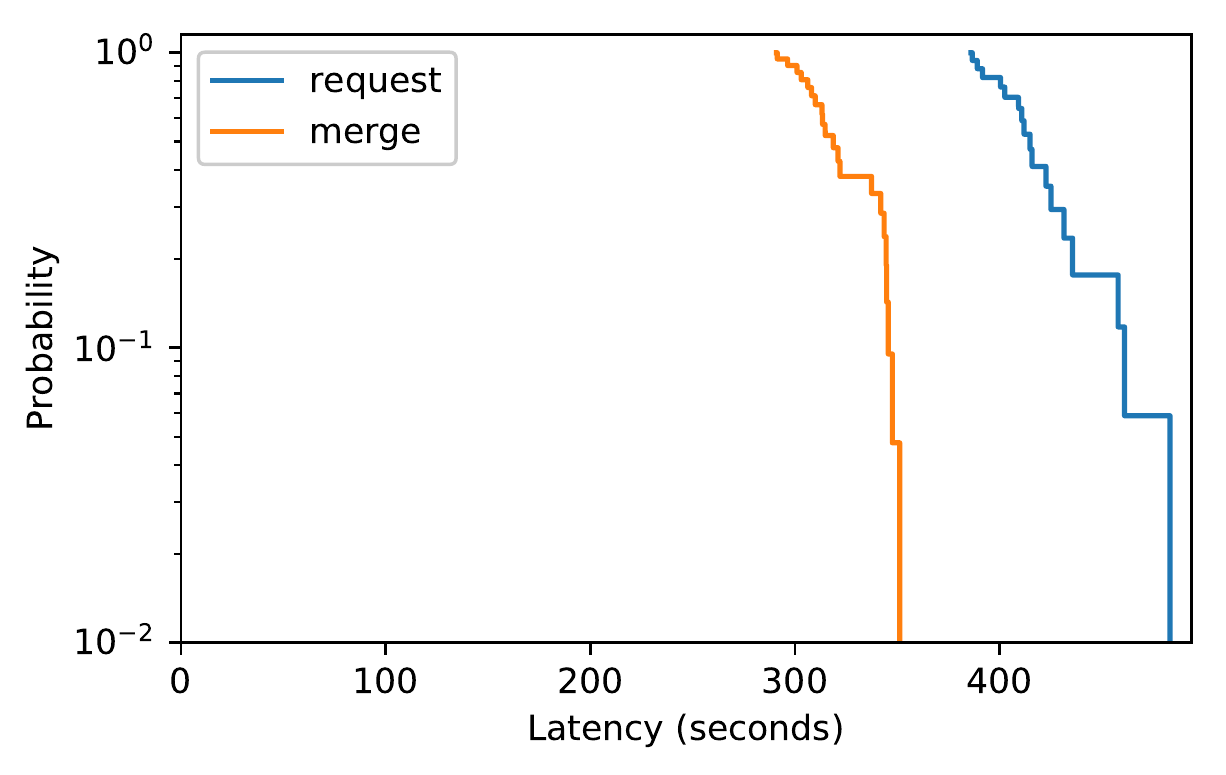}
    \caption{Complementary CDF of component and end-to-end request latencies.
      This configuration uses 3 merge and 17 fused \alignsort local pipelines with 7 open requests.}
    \label{fig:latency_ccdf}
\end{figure}

\quickfig{fig:latency_ccdf} shows complementary CDF of latencies within the fused \alignsort application from an experiment with this maximal configuration.
This figure shows the end-to-end request latency as well as the latencies for the merge local pipelines for a 15-minute duration of the \app operating at a steady state.
Each request requires \dataInputChunksPerRequest align operations, \dataSortsPerRequest sort operations (due to the grouping factor of \dataGroupingFactor in the batching dequeue preceding the sort \stage), and a single merge operation.
The mean latencies for the align, sort, and merge phases is \dataCCDFMeanAlignLatency, \dataCCDFMeanSortLatency, and \dataCCDFMeanMergeLatency, respectively.

\quickfig{fig:latency_ccdf} confirms that (1) \system reduces the serial latency by overlapping different phases of the application across parallel local pipelines; (2) the flow-control and scheduling mechanisms of \system minimize tail latencies well up to the $99^{th}$ percentile; (3) the end-to-end request latency shows greater variability than the component latencies, and is the result of the combined effects of barrier delays and out-of-order \feed delivery of concurrent requests.
Nevertheless, the mean request latency is \dataCCDFMeanRequestLatency but the $99^{th}$ percentile is only \dataCCDFTailRequestLatency.

\subsection{Scale-Out Performance}
\label{sec:eval_persona}

\begin{figure*}[!ht]
  \subfloat[Scale-out throughput as a function of the number of fused align pipelines]{
    \includegraphics[width=0.485\textwidth]{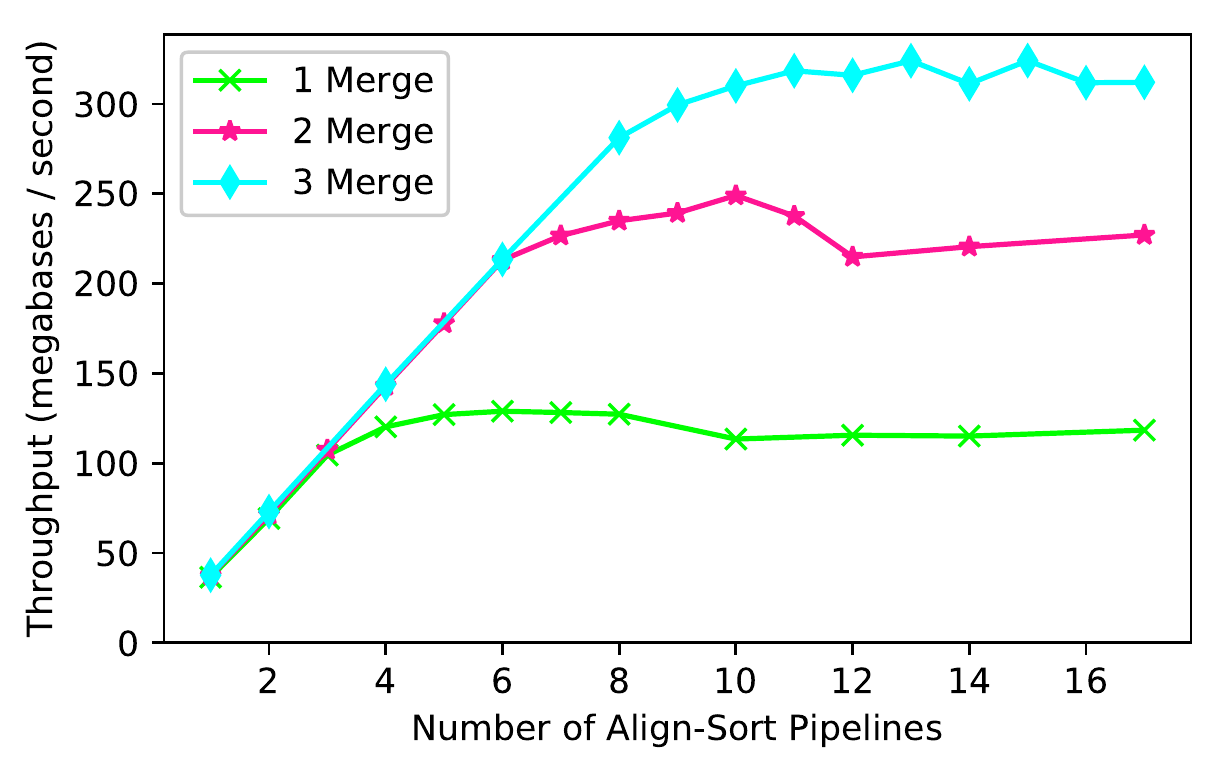}
    \label{fig:fused_scaleout:throughput}
    }
  \hfill
  \subfloat[Scale-out latency as a function of the number of fused align pipelines]{
    \includegraphics[width=0.485\textwidth]{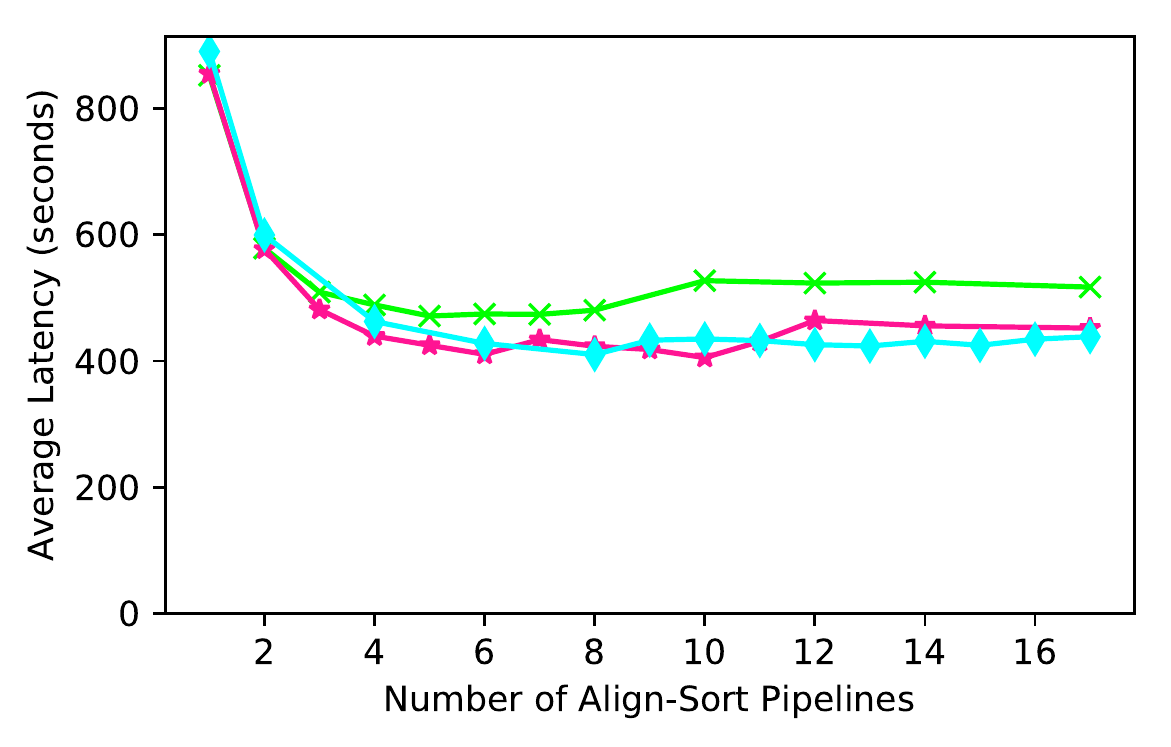}
    \label{fig:fused_scaleout:latency}
    }
  \caption{Scale-out behavior for the fused \alignsort application for 1, 2, and 3 local merge pipelines, configured with 3, 5, and 7 open requests, respectively.}
    \label{fig:fused_scaleout}
\end{figure*}

\quickfig{fig:fused_scaleout} demonstrates that \system can take advantage of additional cluster nodes to both increase the overall throughput and reduce the processing latency of any request processed by the pipeline.
In each series, we fix the number of local merge pipelines and open \batchs (sufficient to saturate the merge pipelines) and scale the number of fused \alignsort local pipelines.

\quickfig{fig:fused_scaleout:throughput} demonstrates the scale-out behavior of the fused \alignsort application's throughput, showing that \system is able to effectively saturate the hardware resources of a balanced pipeline.
The application scales linearly across the hardware resources to the point of saturating one phase of the application.
The results show that approximately 4 aligner nodes saturate one merge node.

\quickfig{fig:fused_scaleout:latency} shows the effects of the scale-out behavior on latency.
As the pipeline can dedicate all aligner nodes in parallel to a single \batch, the latency decreases according to Amdahl's law~\cite{amdahl}, with the remaining latency due to the merge stage.
The latency of a local merge pipeline to merge and compress one dataset is an average of \dataMergeLatency, excluding I/O, for all configurations.

\subsection{Benefits of Fusing Align and Sort}
\label{sec:eval:fused_vs_std}

The fused \alignsort application enables the user to configure fewer machines and incur less I/O compared to the baseline application.
Specifically, the fusion leads to a balanced use of each cluster node's compute and I/O resources, whereas the baseline pipeline has a mix of nodes that are either CPU-bound (the aligners) or I/O-bound (the sort nodes).

Both the fused \alignsort and the baseline's align-only local pipelines are bound by the align \stage.
The sort \stage is relatively inexpensive and takes advantage of the fact that the data is already read and decompressed into buffers to perform the alignment operation: a tuned configuration dedicates 47 aligner threads in the align-only case and 45 threads in the fused \alignsort case.
The minor reduction in throughput of the node is more than compensated by the reduction of \dataFusedVsBaselineDataSaving in aggregate I/O and the elimination of dedicated sort nodes.
Furthermore, fusing aligning and sorting leads to balanced use of CPU and NIC resources of each node of the cluster.

\begin{figure*}[t]
  \subfloat[The aggregate application throughput of both the fused \alignsort and merge phases, averaged over a 5-second window.]{\includegraphics[width=0.49\textwidth]{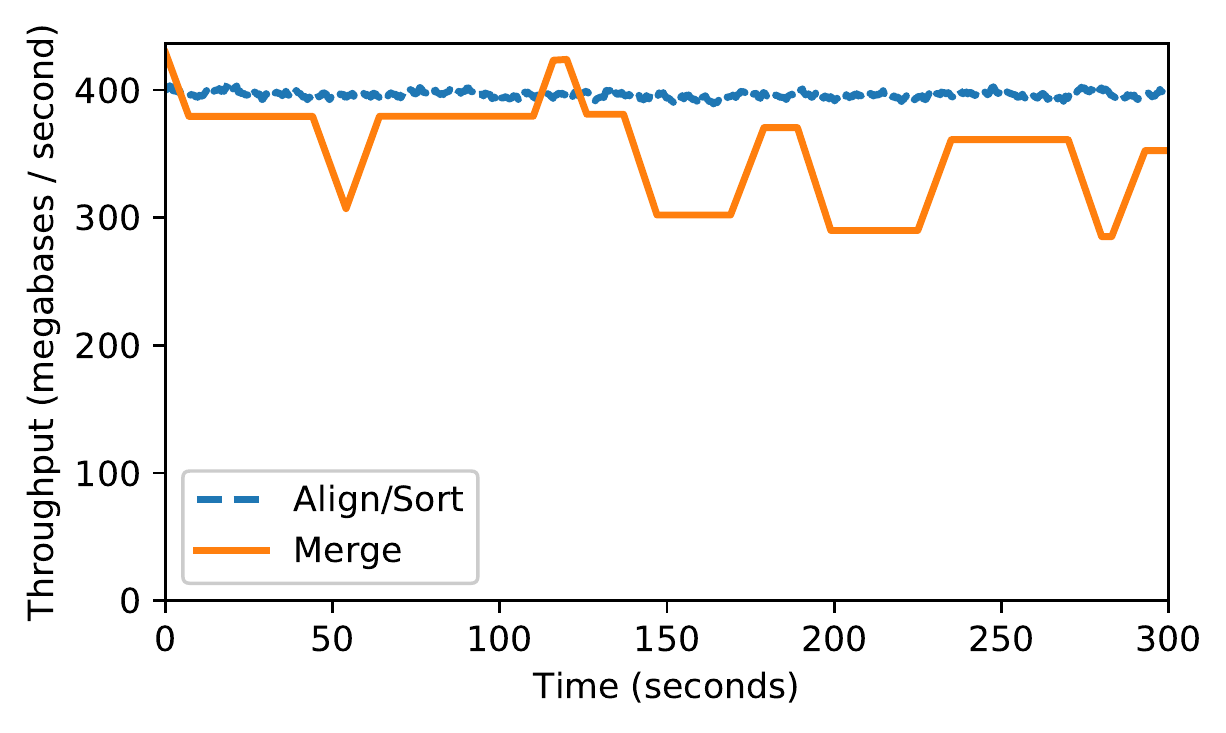}
    \label{fig:fused_runtime:throughput}
  }
  \hfill
  \subfloat[The aggregate I/O behavior for the Ceph read and write \stages of each pipeline. This is the sum of the NIC-level I/O for the read and write \stages aggregated across all of the pipelines, separated by type (merge or fused \alignsort).]{\includegraphics[width=0.49\textwidth]{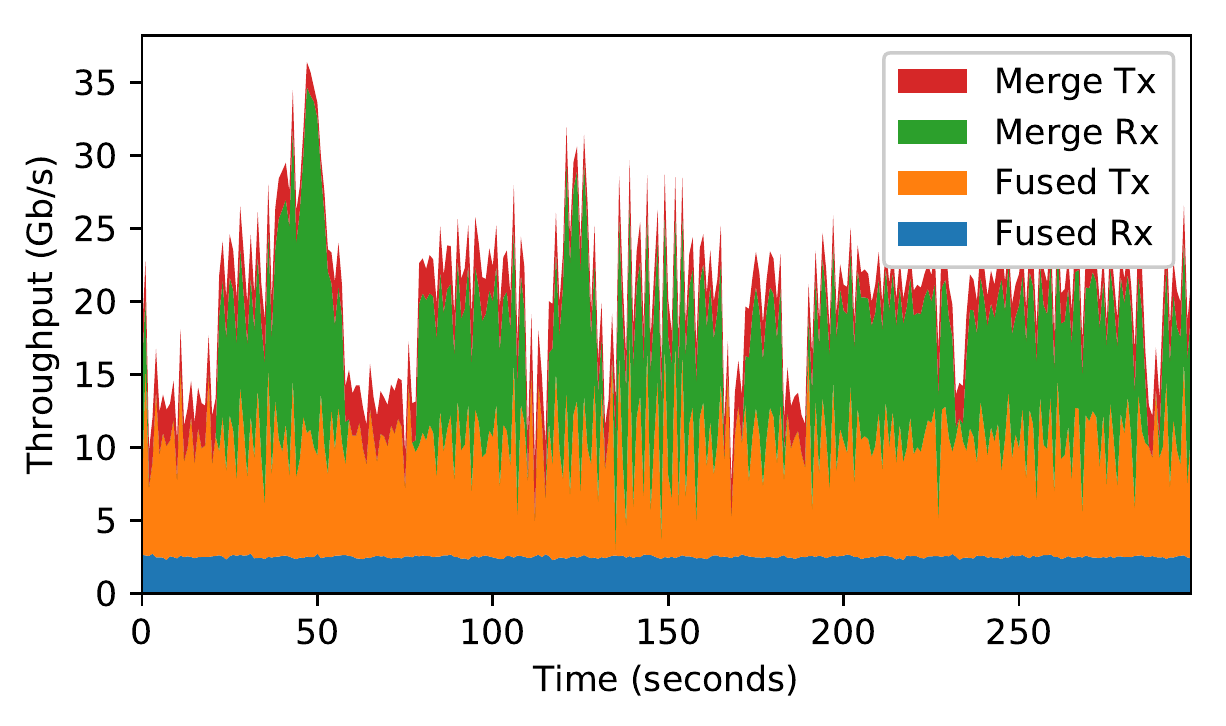}
    \label{fig:fused_runtime:io}
  }
    \caption{The aggregate steady-state behavior of both throughput and I/O for the fused \alignsort application for a period of 5 minutes. The experiment uses 3 merge pipelines, 10 fused \alignsort pipelines, and 7 open requests. This is a maximum-throughput configuration for our cluster.}
    \label{fig:fused_runtime}
\end{figure*}

\quickfig{fig:fused_runtime} shows the steady-state aggregate throughput and I/O of the fused \alignsort application.
These figures demonstrate that \system enables \app to overlap I/O and compute throughout the pipeline.
The I/O rate shown in \quickfig{fig:fused_runtime:io} shows the aggregate I/O rate measured at the NIC; the aggregate is computed across all local pipelines of the same type.
The application throughput shown in \quickfig{fig:fused_runtime:throughput} shows the aggregate throughput of the computationally-intensive parts of the local pipeline.

\section{Discussion}
\label{sec:discussion}

\myparagraph{External vs. Internal Control}
With the addition of metadata to each \feed, \system pipelines perform both data processing and control functionality of a cloud computing framework.
Due to \tfz design decisions outlined in \S\ref{sec:background:tf} and the dataflow rules enforced by the \tf runtime, other cloud computing applications of \tf have typically invoked \tf graphs as a subcomponent of a larger cloud computing framework.
For example, TensorFlowOnSpark~\cite{yahoo-tf-spark} invokes \tf graphs as a subcomponent of a Spark~\cite{spark} application.
This combination of using \tf for processing the data associated with incoming \batchs while delegating scheduling and controller responsibilities to an external framework is pragmatic, but it comes at a high cost: transforming data to and from \tfz internal data representation is an expensive operation, dominating the runtime of numerically intensive applications such as machine learning~\cite{weld}.
By attaching metadata to each \feed with minimal modifications to \tf (and none to the core \tf runtime), \system can process each \batch by itself, without relying on an external framework.

\myparagraph{Parameter Tuning} 
Tuning the parameters in a \system pipeline is important to maximize application throughput while avoiding excessive resource usage.
A properly configured pipeline will be bound by the throughput of a hardware resource \egp{the CPU required for the align \stage of \app} while avoiding excessive resource usage;
excessive resource usage occurs when a \stage and \gate upstream of the hardware-bound \stage in the pipeline is not bounded, \eg if excess buffers are used in a read or decompression \stage upstream of the align \stage.
The key parameters to tune are
(1) the number of \stages within a local pipeline,
(2) the number of local pipelines to replicate for each phase (for \app: align, sort, \alignsort, and merge)
(3) the number of open \batchs to allow within a local or global pipeline (limited by one or more credit links).
We observe the state of an application using hardware statistics \egp{CPU and memory usage per local pipeline}, tracing to record events during an execution of an application \egp{when a read or write \stage executes}, and the capacity and buffer size of each \gate using the Tensorboard mechanisms from \tf.
In practice, one scales the \stages in a local pipeline until a hardware resource is saturated and then balances the number of machines allocated to each local pipeline such that the aggregate throughputs of all local pipelines are approximately equal.

\myparagraph{Compatibility with \tf}
In practice, a \tf application consists of the interaction between a declarative graph and a client program.
The \tf client program of a typical supervised machine-learning pattern repeatedly feeds training examples to the graph and uses gradient descent to backpropagate updates to the model.
The algorithm terminates after prediction error falls below a threshold or a fixed number of training examples are processed.

While \system does not have the concept of a client program, \system \gates provide a mechanism to encode commonly-used patterns within the \system pipeline.
\system supports supervised training of a machine learning model, feeding multiple \batchs of training examples against the model within the pipeline.
New \gates can extend \systemz functionality to support this use case (see the aggregate dequeue operation in \S\ref{sec:arch:gates}).
We leave the evaluation of such gates to future work.

\myparagraph{Comparison to \tf loops}
\tf has a loop construct~\cite{tf-loops}, which can be used to construct an execution pipeline that iterates across an aggregate \feed containing all \feeds of a given \batch.
While this allows concurrent execution and I/O overlap, \batchs must be statically partitioned across machines.
\system enables dynamic workload partitioning, provides a built-in flow-control mechanism that limits resource usage, and allows for the online submission of a stream of \batchs.
In addition, \tf loops can be naturally used within the \stage graph of a \system pipeline, \eg to dynamically change the computation based on the value within a \feed.


\eat{
the \stage logic graph can be performed in the loop body iterating across an aggregate of all \feeds in a single \batch, and an entire pipeline can be expressed as a linear sequence of loop nodes that apply successive \stage logic to this aggregate \feed.
Provided this pipeline uses no \tf queues, it is possible to concurrently process multiple \batchs on the same application instance.
\systemz architecture improves on this approach by decoupling \batch semantics from application semantics: whereas the loop approach imposes a barrier for each \batch between every \stage, \system enables individual \feeds from each \batch to be pipelined across multiple successive \stages for concurrent processing.
Without \gates to distribute \feeds to local pipelines, a loop-based approach must statically partition each \batch, thereby increasing the latency due to the work imbalance between partitions.
\systemz use of \gates enables \feed processing and latency to be decoupled from the structure of the application graph by distributing \feeds to each local pipeline or \stage based on \feed availability and FCFS semantics.
}

\myparagraph{Fault tolerance}
\ptf inherits \tfz fault tolerance semantics because it does not alter the \tf runtime.
\tf does not implement fine-grained recovery, instead relying on coarse-grained checkpointing to periodically save the state of training models;
\tf does not include \feed-level fault recovery due to the high overhead.
\ptf currently relies on the exactly-once semantics of \feed delivery from \tf;
it can relax this requirement to at-least-once semantics with additional components added to the metadata.
For example, an additional \feed ID can be inserted into the existing metadata tensor to differentiate each \feed within a \batch, effectively creating a compound ID (for both the \batch and the \feed) that uniquely identifies the \feed between any pair of adjacent \gates.
\ptf will inherit any \feed-level recovery mechanism from \tf; efforts in this regard are orthogonal and complementary to \ptf.

\section{Related Work}
\label{sec:related_work}

\system builds on a multitude of related work.

\myparagraph{Dataflow} \systemz use of metadata to differentiate \feeds of concurrent \batchs is inspired by the tagged token dataflow (TTDF) architecture~\cite{tagged-dataflow}.
\tf incorporates many principles from TTDF, primarily using a runtime tag to associate related tensors of a given \feed when concurrently processing multiple \feeds.
\system uses an explicit tag, \ie metadata, to relate different \feeds across successive invocations of a graph.

\myparagraph{Persona} \app builds upon Persona \cite{persona}, which consists of both a library of \tf nodes for bioinformatics processing and a corresponding application (written in Python) that uses the library code to build applications that process a single request.
\system is independent of Persona; \app depends on both \system and Persona.
We use the library components of Persona in developing \app \egp{I/O, alignment, compression}, but rewrite \app as a new application around \system due to the single-request design of its applications.
This is due to the differing goals of Persona and this work: Persona creates high-performance bioinformatics applications while \system creates high-performance cloud computing streaming applications.
We use Persona's library to combine these two goals by developing \app as an application for streaming bioinformatics pipelines that are capable of running as persistent services using \systemz unique features.

\myparagraph{Batching cloud computing frameworks} MapReduce \cite{mapreduce} is the first modern BSP-style \cite{bsp} framework to be used for cloud computing applications.
Dryad \cite{dryad} extends this architecture to support more general-purpose dataflow applications.
Spark \cite{spark} improves the performance of these operations by introducing the Resilient Distributed Dataset, which incrementally caches datasets in memory across machines.
All frameworks use a coordinator to schedule tasks between distinct stages.
This coordination provides a convenient location to apply cluster scheduling and fault recovery logic.
\system eschews this model, as the scheduling overhead for multiple tasks imposes a higher latency cost for features not targeted by this work:
scientific applications only require coarse grain recovery mechanisms \iep{restarting the application} as they typically are append-only for data provenance.

\myparagraph{Streaming cloud computing frameworks} This category of frameworks encode their logic as a dataflow graph, with each node being a long-running process that reacts to events, such as the arrival of new data on an input edge, by producing new events, such as outputting new data on an output edge.
Flink \cite{flink}, StreamScope \cite{streamscope}, and Millwheel \cite{millwheel} are recent examples of this architecture.
Naiad \cite{naiad} extends the streaming model to include more general graphs than direct acyclic graphs, but at the cost of higher communication overhead.
\system adopts this model for execution due to its low scheduling overhead and encodes the notion of punctuations \cite{streaming-punctuations} with the metadata signaling a punctuation when a request has been fully processed by a \gate.

\myparagraph{\tfz use in the cloud computing ecosystem}
Several research initiatives seek to allow \tf to be used as a subcomponent of a larger application in a cloud computing ecosystem.
Weld~\cite{weld} unifies multiple cloud computing frameworks into a single runtime data representation, providing adapters for many existing frameworks including \tf.
Apache Arrow~\cite{apache-arrow}, a columnar data format for inter-framework interoperability, was recently added to Spark~\cite{spark-arrow} for a similar reason.
This can be combined with TensorFlowOnSpark \cite{yahoo-tf-spark} to distribute training via Spark RDDs with low overhead \cite{tf-spark-arrow}.
These efforts are complementary to \system: interoperability with other frameworks with lower overhead benefits all.
\eat{
However, this effort does not benefit all applications; Persona relies on custom data structures for efficiency \egp{memory-mapped file handles} and only passes handles to those resources between \tf nodes.
\system is implemented entirely in \tf to avoid copying data to and from an external cloud computing framework while remaining compatible with existing \tf applications such as Persona.
}

\myparagraph{Native execution} Nimbus \cite{nimbus} overcomes the limitations imposed by JVM-based frameworks by implementing a native dataflow engine in C++ and caching scheduling decisions through the use of execution templates.
Both Nimbus and \system approach the problem of CPU-bound cloud computing tasks by implementing an alternative to the overheads imposed by the JVM.
Whereas Nimbus combines the application logic and runtime into a single binary for the worker nodes in its architecture, \system decouples the runtime and application logic;
instead of distributing a binary with the application logic, \system sends only a serialized description of the application graph to each machine.

\section{Conclusion}
\label{sec:conclusion}

This work demonstrates that by introducing a few careful abstractions into \tf \iep{\stages and \gates}, \system can operate as a framework for constructing a wider variety of cloud computing applications and running these applications on a scale-out cluster of machines.
\system augments \tfz existing runtime and library of nodes to enable the creation of scientific cloud computing applications that operate as a service, running indefinitely while concurrently processing multiple requests.
\system serves as both the data processing and controller components of such a framework without relying on an external agent to coordinate computation: \stages partition the application's graph into independent components for processing, \gates control and coordinate the progress of concurrent requests between \stages, and pipelines describe sequences of \stages and \gates that span multiple machines.

We implement the extension in \system and use it to develop \app, a streaming bioinformatics pipeline built with components from Persona.
When scaled across \clusterSize machines, \app is able to align and sort \dataMaxTP while pipelining \dataMaxTPrequiredOpenRequests open requests in parallel.
We use \system to partition and join requests between global and local pipelines to develop a fused version of the standard align-sort-merge pipeline, eliminating an I/O phase between alignment and the first phase of sorting by combining both pipelines on a single machine.
This saves a total of \dataFusedVsBaselineDataSaving of data transfers to the storage subsystem.

\newpage
\bibliographystyle{acm}
\bibliography{gen-abbrev,dblp,references}

\end{document}